\documentclass[a4paper,fleqn]{cas-sc}

\usepackage[authoryear,longnamesfirst]{natbib}

\usepackage{subfigure}
\usepackage{lineno}
\usepackage{comment}
\usepackage{color}

\usepackage{multirow}
\usepackage{booktabs}

\usepackage{adjustbox}
\usepackage{rotating}
\usepackage{makecell}

\usepackage[linesnumbered,ruled,vlined]{algorithm2e}

\usepackage{pifont}

\def\tsc#1{\csdef{#1}{\textsc{\lowercase{#1}}\xspace}}
\tsc{WGM}
\tsc{QE}
\tsc{EP}
\tsc{PMS}
\tsc{BEC}
\tsc{DE}

\begin{document}
\let\WriteBookmarks\relax
\def\floatpagepagefraction{1}
\def\textpagefraction{.001}

\shorttitle{Bibliometris on Authentication and Threat Model}

\shortauthors{Bezerra, Souza, Westphall and Westphall}

\title [mode = title]{A Bibliometrics Analysis on 28 years of Authentication and Threat Model Area}              



%
\author[1,3]{WR Bezerra}[
                        role=Phd Candidate, orcid=0000-0002-6098-7172]
\cormark[2]
\fnmark[1]
\ead{wesleybez@gmail.com}

\author[1,3]{CA de Souza}[
                        role=Phd Candidate, orcid=0000-0002-9453-3240]
\cormark[2]
\fnmark[1]
\ead{cristianoantonio.souza10@gmail.com}

\author[1,3]{CM Westphall}[
                        role=Phd, orcid= 0000-0002-7213-1603]
\cormark[2]
\fnmark[1]
\ead{carla.merkle.westphall@ufsc.br}

\author[1]{Carlos B. Westphall}[role=Advisor,orcid=0002-5391-7942]
\cormark[1]
\ead{carlosbwestphall@gmail.com}

\address[1]{UFSC - Federal University of Santa Catarina,Campus Universitário - Trindade,Florianópolis/SC - Brazil, 88040-380}

\cortext[cor1]{Corresponding author}
\cortext[cor2]{Principal corresponding author}



\begin{abstract}
   The large volume of publications in any research area can make it difficult for researchers to track their research areas' trends, challenges, and characteristics. Bibliometrics solves this problem by bringing statistical tools to help the analysis of selected publications from an online database. Although there are different works in security, our study aims to fill the bibliometric gap in the authentication and threat model area. As a result, a description of the dataset obtained, an overview of some selected variables, and an analysis of the ten most cited articles in this selected dataset is presented, which brings together publications from the last 28 years in these areas combined.
\end{abstract}


\begin{highlights}
\item obtaining a sufficiently representative dataset of the authentication and threat model area to support future research.

\item a set of statistical analyzes of publication data, such as sources, publication by time, and most cited publications.

\item analysis of the top ten articles with the highest number of citations of the entire period and classification of these works in categories.
\end{highlights}

\begin{keywords}
threat model \sep authentication \sep security \sep bibliometrics
\end{keywords}

\maketitle

\section{Introduction}
\label{S:1}

Robust analysis is based on data and instruments applied to obtain these data. Regarding it, this study analyzed the results of the database gathered in scientific publications portals using bibliometrics instruments. Such instruments were part of the threat model proposal for a Ph.D. thesis; however, it is not limited to this purpose and can also provide a bibliometric basis for any studies involving authentication and threat model.

Bibliometrics, or bibliometric analysis, is the method that allows us to analyze a large number of scientific studies in an objective way \cite{wang2021evolution}. Therefore, it allows analyzing statistical data and research trends within its specific areas. For example, it is possible to assess the growth in the number of publications on vaccines and viral diseases after the emergence of COVID-19 at the beginning of 2020 in Brazil.

In academia, researchers should make decisions based on metrics that are more accurate and complete \cite{wilsdon2015metric}, being bibliometrics an essential discipline for research. This term was first used by Alan Pritchard in 1969 \cite{broadus1987toward} and can be defined as using statistics in the service of pattern analysis in publications \cite{mcburney2002bibliometrics}. Still, the terms Bibliometrics, Scientometrics, and Informetrics can be confused due to their similarities in application \cite{hood2001literature}; however, there are some differences even though most of the works classified as bibliometrics are informetrics. Informetrics mainly uses digital media and electronic databases to perform \cite{hood2003informetric} statistical analysis.

\begin{table}[h!]
    \centering
    \caption{Related Work}
    \begin{adjustbox}{max width=\textwidth}
    \begin{tabular}{@{}p{.2\linewidth}lll@{}}
    \toprule
        Article & Category & Area & Content\\
        \toprule
        \cite{dayyabubibliometric} & Security & Cloud & \makecell[l]{Brought a work in the large area of security, specifically for cloud computing service,\\ presenting a bibliometric analysis of the topic.} \\
        \midrule
        \cite{farooqui2021bibliometric} & Security & 5G & \makecell[l]{Security challenges and trends in the 5G area are presented.}\\
        \midrule
        \cite{pathak2021bibliometric} & Security & Authentication & \makecell[l]{Presents bibliometric research on Zero Knowledge Proof (ZKP) for Authentication, using\\ a dataset of 329 articles from 2010 to 2021, the authors demonstrate that it has  relevant\\ for countries like China, the USA, and India.}\\
        \midrule
        \cite{rejeb2021blockchain} & Blockchain & Smart Cities & \makecell[l]{Presents an analysis of co-occurrence of keywords and co-citation to demonstrate the\\ growing use of this technology in Smart City and list its trends. }\\
        \midrule
        \cite{pandey2022blockchain} & Blockchain & Supply Chain & \makecell[l]{Brings another work on blockchain, however, applied to the supply chain and discussed\\ its trends, main researchers, and in which areas of application their research are\\ concentrated.}\\
        \midrule
        \cite{azan2022blockchain} & Blockchain & Supply Chain & \makecell[l]{Specifically in traceability, this work presents a bibliometric and lexicometric analysis\\ to obtain the chronology and evolution of these two areas.}\\
        \midrule
        \cite{maduri2022bibliometric} & Blockchain & Insurance & \makecell[l]{Brings an analysis of blockchain applications in the insurance industry with the purposes\\ of mitigating cyber-threats, increasing customer trust, and reducing costs.}\\
        \midrule
        \cite{kumar2022artificial} & Blockchain & Artificial Intelligence & \makecell[l]{Presents an analysis on blockchain and artificial intelligence finding ten main areas of\\ application in business of this technology, from an analysis that obtained four clusters\\ with a focus on supply chain, healthcare, secure transactions, and finance and account. }\\
        \midrule
        \cite{sakhnini2021security} & - & Smart Grids & \makecell[l]{Brings a bibliometric survey presenting gaps and threats regarding the security of using\\ IoT in Smart Grids}\\
        \midrule
        \cite{rani2022bibliometric} & - & Smart Cities & \makecell[l]{Brings an analysis of research on actuators used for automation of services in Smart\\ Cities.}\\
        \bottomrule
    \end{tabular}
    \end{adjustbox}
    \label{tab:related_work}
\end{table}

Regarding the area of security, threats, and authentication, several works stand out and present important results for the area (Table \ref{tab:related_work}). However, we grouped the works to provide a overview. Firstly, the more general works present applications in several areas such as Supply Chain \cite{pandey2022blockchain,azan2022blockchain}, Smart Cities \cite{rejeb2021blockchain,rani2022bibliometric}, Smart Grids \cite{sakhnini2021security}, or involving other technologies such as artificial intelligence \cite{kumar2022artificial}, among others \cite{maduri2022bibliometric}. Others are more focused on security, such as security in the cloud \cite{dayyabubibliometric}, in 5G \cite{farooqui2021bibliometric}, and authentication \cite{pathak2021bibliometric}. Besides, although there are several interesting works on bibliometrics in the area, our work is paramount to addressing specific issues of threat models and authentication that are necessary for research in the area.

The bibliometric analysis method brought statistical evidence to analyze documents related to authentication and threat models in a broader scope. More specifically, this work aims to answer the following urgent questions:
\begin{itemize}
    \item[Q1] what is the \textbf{overview of threat model and authentication} provided by bibliometrics?
    \item[Q2] what is the \textbf{main publishing sources} in the area?
    \item[Q3] what are the \textbf{main works} in the area?
\end{itemize}

The work follows the following structure: Section \ref{S:2} presents the methodology steps. Hence, the Subsection \ref{S:2.1} describes the data sources used. Also, the Subsection \ref{S:2.2} overviewed the study's characteristics. Still, in the Subsection \ref{S:2.3}, the ten most cited articles are commented on. Therefore, Section \ref{S:3} shows results and discussion on this work. Finally, Section \ref{S:4} brought this study's conclusion and future works.

\begin{table}[]
    \centering
    \begin{tabular}{ll}
    \toprule
        Abbreviation & Meaning  \\
        \toprule
        CPS & Cyber-Physical System\\
        WoS & Web of Science\\
        ZKP & Zero Knowledge Proof\\
        \bottomrule
    \end{tabular}
    \caption{List of Abbreviations}
    \label{tab:my_label}
\end{table}

\section{Bibliometrics Analyses}
\label{S:2}

Regarding methodology, from the perspective of a data-based analysis, the present study analyzes the relationship between "threat model" and authentication through bibliometrics. 
The works are limited to articles from reviews, conferences, and journals. As a result, a database in BibTeX format was obtained, which can be viewed at the following link\footnote{https://github.com/wesleybez/mfar\_tm/databases/}.

Concerning materials, as a tool to support the statistical analysis, RStudio\footnote{https://www.rstudio.com/} was used together with the bibliometrix package (biblioshiny)\footnote{https://www.bibliometrix.org/}. RStudio is a shell that encapsulates the R language interpreter (r-lang) and allows the visualization and editing of data. Also, this tool allows editing and executing scripts in r-lang. In turn, bibliometrix is a package that extends the functionality of r-lang, incorporating functions for analyzing bibliometric data, correlations between bibliometric variables, and importing data sources from research portals, such as the Web of Science (WoS) \footnote{https://www-webofscience.ez46.periodicos.capes.gov.br/wos/scielo/advanced-search} and Scopus\footnote{https://www.scopus.com}. All statistical analysis and generation of graphs and tables happened with the support of this tool. The scripts used can be found in the repository of this publication\footnote{https://github.com/wesleybez/mfar\_tm}.

As for the query used, the query-string was submitted to the Scopus platform on September 24, 2021, and returned several documents organized in a database for later analyses. Although the number is not so expressive, it served as an initial entry for the bibliometrix. The number of studies obtained allowed us to analyze factors such as:
\begin{itemize}
    \item annual scientific production trend;
    \item average number of citations per year;
    \item main publishing sources;
    \item top ten most cited articles.
\end{itemize}

Each cited item is evaluated in detail throughout the text, allowing an overview and an introduction to the data extracted from the study. Still, such data allow future works can concisely analyze the data sources and extract more statistical analysis from them according to the needs of the proposed works. Next, the data sources (\ref{S:2.1}) are analyzed.

\subsection{Data Sources Description}
\label{S:2.1}
Regarding the general analysis of the data sources, it briefly presented the data collected during the research. Thus, such a dataset can be better analyzed in the present study and in terms of providing artifacts for future research. Let us start with the chosen time window, the time interval analyzed.

As a time window, a \textbf{time interval} ranging from 1994 to 2022 (28 years) was analyzed. Further, in this interval, 922 documents appear in 550 sources of different publications. Since, these documents together showed a total of 46208 references that supported these works. Therefore, this dataset is representative and comprises an adequate time interval for the present study.

Specifically, about the \textbf{types of documents} analyzed, a total of 333 articles appear in the results (36.11\%), nine books (0.09\%), and 35 book chapters (3.79\%) are present in the study. As for publications of the "conference papers" type, they appear with a large number of works, a total of 493 works (53.47\%), followed by reviews with 41 occurrences (4.44\%), and in third place the "conference reviews" with seven papers (0.75\%). The results show that articles in conferences are the type of publication that most addresses the researched topic.

Regarding the \textbf{annual distribution of publications}, they started in 1994 and gained a major proportion of growth from 2005 onwards. However, the largest volume of publications occurs from the last ten years ($>=$ 2012), with a total of 762 publications (82.64\%). Further, it is worth emphasizing that this growth continued and increased in the last five years, bringing a total of 534 documents (57.91\%) from the entire series. Therefore, it is observed that this description does not consider the year 2022 in the segmentation mentioned before. Consequently, it shows that these areas are important research topics today.

Concerning the \textbf{authors}, they form a complete set of 2,374 individuals. 
Concerning this number, 66 published a single-authored document and 2,308 published works together with other authors; approximately 97\% of the works were developed and published in partnership between the authors. In brief, the collaboration between authors is shown to be one of the main factors for the evolution of science in different groups and institutions. Thus, we note that collaboration between authors is common in the area.

Furthermore, as \textbf{complementary indicators} on the relationship between authors and publication, there is a total of 0.38 documents for each studied author. However, when analyzing the inverse relationship, a total of 2.57 authors are obtained for each document, on average. Through this data, there is an indication that few authors remain researching and reporting in the area.

As for \textbf{authors' productivity}, the authors with the highest number of published documents and tied for first place are professors Mohamed Amine Ferrag and Sudeep Tanwar, with 11 works. Second, there is a triple tie, all authors with ten publications in the results, between professors Giampaolo Bella, Neeraj Kumar, and Jingyue Li. Thereupon, this difference of one unit between first and second place demonstrates how the cited authors have an equally significant contribution to the research area regardless of rank.

In short, it is concluded that the data set has a representative distribution of the items evaluated. Note that the time window, types of documents, authors, complementary, and productivity were briefly analyzed in this section. In this way, these chosen variables present paramount importance in analyzing the chosen topic, its publications, and its relevance as a current and innovative research topic.

\subsection{General Analysis of Characteristics}
\label{S:2.2}

This subsection presents a general analysis of the listed characteristics in this study. An overview was drawn from the statistical data allowing us to better understand the main sources, publications dynamics, and the articles analyzed. As a result, this section commented the threat model and authentication research contained in 28 years of publications.

\subsubsection{Source Overview}
\label{S:2.2.1}

Publication sources are one of the most important information about articles. The author can draw up publication plans from this information, monitor journals, and follow the research topic evolves in different sources. In addition, the analysis of the most important sources for a research topic allows the researcher to focus efforts on more specialized sources in his/her topic and to get to know possible journals and events that can contribute to the evolution of such work through the publication and reviewing. In brief, this work presented the most relevant publication sources found for the topic of authentication and threat modeling.

\begin{table}[h!]
    \centering
    \caption{Most Relevant Sources - in the first column indexing is presented, the second column brings the name of the source, in the third column, the total of articles is presented, and finally, the categorization (Proceeding/Journal)}
    \begin{tabular}{@{}llll@{}}
    \toprule
    \#&Source&Articles&Category\\
    \midrule
    s1&  \makecell[l]{LECTURE NOTES IN COMPUTER SCIENCE (INCLUDING\\ SUBSERIES LECTURE NOTES IN ARTIFICIAL INTELLIGENCE\\ AND LECTURE NOTES IN BIOINFORMATICS)} & 77& PROCEEDING\\
    \midrule
    s2&  \makecell[l]{ACM INTERNATIONAL CONFERENCE PROCEEDING SERIES} & 19& PROCEEDING\\
    \midrule
    s3&  IEEE ACCESS & 19& PROCEEDING\\
    \midrule
    s4&  \makecell[l]{PROCEEDINGS OF THE ACM CONFERENCE ON COMPUTER AND\\ COMMUNICATIONS SECURITY} & 16& PROCEEDING\\
    \midrule
    s5&  \makecell[l]{COMMUNICATIONS IN COMPUTER AND INFORMATION SCIENCE} & 10& PROCEEDING\\
    \midrule
    s6&  COMPUTER NETWORKS & 10&JOURNAL\\
    \midrule
    s7&  \makecell[l]{IEEE INTERNET OF THINGS JOURNAL} & 09&JOURNAL\\
    \midrule
    s8&  COMPUTERS AND SECURITY & 08&JOURNAL\\
    \midrule
    s9&  \makecell[l]{CONFERENCE ON HUMAN FACTORS IN COMPUTING SYSTEMS\\ - PROCEEDINGS} & 07& PROCEEDING\\
    \midrule
    s10& \makecell[l]{IEEE COMMUNICATIONS SURVEYS AND TUTORIALS}  &  06&JOURNAL\\
    \bottomrule
    \end{tabular}
    \label{tab:most_relevant_sources}
\end{table}

Specifically regarding \textbf{publication sources} (Table \ref{tab:most_relevant_sources}), we can divide them into "proceedings" and journals. In the list of top ten sources, the first five are proceedings, and of these, Springer's Lecture Notes in Computer Science(LNCS) comes first, with 77 publications. Followed by the ACM International Conference Proceeding Series, with a much lower value of 19 publications (approximately 25\% of the first place).

Focusing on journals, Computer Networks comes in first with ten publications, followed by the IEEE Internet of Things Journal, with nine publications. The sources cited before are respectively from Elsevier and IEEE. Also, it is observed that although the IEEE Internet of Things Journal came in second place, it exists from seven years (launched in 2014), while Computer Networks had its beginnings in 1977 (44 years) - suffering different periods of publication interruptions in the meanwhile.

At the end of the list, we have six articles from the IEEE Communications Surveys and Tutorials. Although, even though not mentioned, all the top ten sources are of major relevance as data sources for researchers in the threat model and authentication fields.

In summary, assuredly the most cited sources are "proceedings." However, both Elsevier and IEEE present relevant sources of publications in journals that present themselves as relevant even when compared with the proceedings. Among such periodicals, we can highlight the Journal of Internet of Things, from IEEE; and Elsevier's Computers and Security Journal.

\subsubsection{Publications Overview}
\label{S:2.2.2}

Regarding publications, it was analyzed from two perspectives: production by area and the average number of citations per year. Firstly, the production allows analyzing the number of documents produced and their peaks and the trend regarding publications within the chosen research topic. Secondly, the average citation per year, on the other hand, allows for evaluating the results of such publications within the academic community and how these are expressed through the use and citation of these works by other researchers.

\begin{figure}[h!]
    \centering
    \includegraphics[width=.65\textwidth]{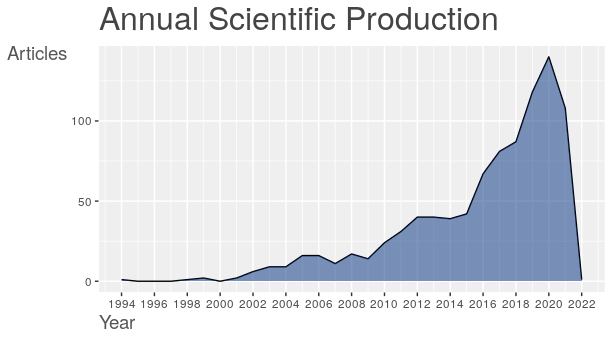}
    \caption{Annual scientific production in the area. In this graph, is noted that the growth starts moderately from 2008 onwards and gains greater proportion in the last seven years.}
    \label{fig:annual_production}
\end{figure}

Focusing on the \textbf{evolution of scientific production} in the area (Figure \ref{fig:annual_production}), it is noted that the largest volume of publications is contained in the last ten years (approximately 82\%). However, if the time window were reduced to five years, the percentage would be approximately 57\%. Such facts demonstrate that research involving threat models and authentication is an area of significant growth. As a result, we can conclude that this growth has gained strength in recent years and continues to increase.

Nevertheless, on the same graph, it can be noted that in three moments, there is a significant increase in the number of publications: first in \textbf{2008}, the second in \textbf{2015}, and the third in \textbf{2018}. The first growth in \textbf{2008} goes until 2012, establishing almost 50 annual publications in the area. The second growth, after a slight decline from \textbf{2015} to 2018, shows a growth each year, expressing the growth of this research area. Finally, the peak from \textbf{2018} onwards confirms this fast-growing trend. As a result, it can be inferred that there is a growth in the production of scientific articles involving threat models and authentication throughout the analyzed period.

\begin{figure}[h!]
    \centering
    \includegraphics[width=.65\textwidth]{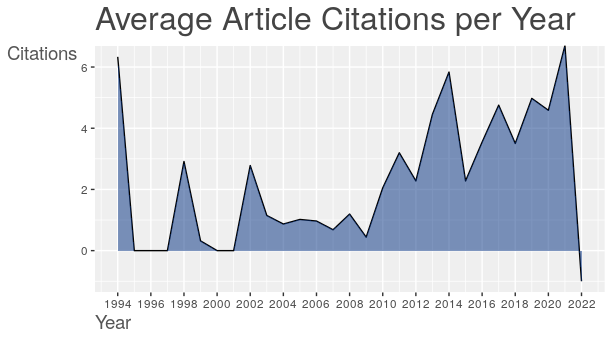}
    \caption{Average citations per year. This graph plots the average citations per year from 1994 to 2022. It is possible to note the growth in the number of citations from 2008 onwards.}
    \label{fig:average_year_citation}
\end{figure}

As for the \textbf{average number of citations per year} (Figure \ref{fig:average_year_citation}), there is a more equally distribution of works. Although a single fundamental work for the area pulls the second biggest peak of the chart to 1994 \cite{anderson1993cryptosystems}, the highest peak of citations takes place in 2020 due to the growth of the research area. 

In this way, it is concluded that it is a topic that has generated important results in a short time. Nevertheless, it is still important to comment that its most recent growth peak demonstrates a tendency to take a more careful look at security issues, especially an authentication and threat model area.

\subsubsection{Articles Overview}
\label{S:2.2.3}

The most cited articles are presented as guides for conducting the state-of-art analysis, and their citations, the articles can be fruitful\footnote{It was confirmed by Forwarding Snow Balling process}, complementing future work, filling in gaps, or even being used as a common source among researchers in the area to deepen their works. Thus, it was necessary to analyze the top ten most cited articles.

\begin{table}[h!]
    \centering
    \caption{Top 10 most cited documents in bibliometric analysis. The documents are sorted in descending order in terms of the number of citations.}
    \begin{adjustbox}{max width=.85\textwidth}
    \begin{tabular}{@{}rp{.2\textwidth}ll@{}}
        \toprule
        \# &Ref.& Title & Citation \\
        \midrule
        a01 &\cite{belanger2011privacy}& \makecell[l]{Privacy in the digital age: a review of\\ information privacy research in information systems } & 658\\
        \midrule
        a02 & \cite{xiao2012security}& Security and privacy in cloud computing & 375\\
        \midrule
        a03 &\cite{bhunia2014hardware}& Hardware Trojan attacks: Threat analysis and countermeasures  & 365 \\
        \midrule
        a04 &\cite{rostami2014primer}& A primer on hardware security: Models, methods, and metrics  & 331 \\
        \midrule
        a05 &\cite{humayed2017cyber}& Cyber-physical systems security—A survey  & 323\\
        \midrule
        a06 &\cite{carlini2016hidden}& Hidden voice commands  & 242 \\
        \midrule
        a07 &\cite{cheminod2012review}& Review of security issues in industrial networks &234\\
        \midrule
        a08 &\cite{bolle2002biometric}& Biometric perils and patches & 233\\
        \midrule
        a09 &\cite{xiao2016hardware}& Hardware trojans: Lessons learned after one decade of research  & 199\\
        \midrule
        a10 &\cite{yan2012security}& Security challenges in vehicular cloud computing  & 184 \\
        \bottomrule
    \end{tabular}
    \end{adjustbox}
    \label{tab:top10_docs_citation}
\end{table}

The analysis of the ten most cited works in the research area, Table \ref{tab:top10_docs_citation}. Firstly, there is the work \textbf{a01} \cite{belanger2011privacy} by Berlanger and Crossler from 2011 with an accumulated 658 citations, followed by the work \textbf{a02} \cite{xiao2012security} by Xiao and Xiao from 2012 with 375 citations. Also, it is important to mention that the oldest work that appears in the list is from 2002, work \textbf{a08} by Bolle \textit{et al.}. with 233 citations. From an overview, it can be said that two moments are very apparent: 2002 with the publication \textbf{a08} and the others organized from 2011 to 2016. In this way, we can infer that the largest volume of citations is from articles from 2011 to 2016, the publications that form the bibliographic base of current works.

\subsection{Analysis of the Top Ten Articles}
\label{S:2.3}

In this subsection, the contributions of each article in the list of the ten most cited were presented in this analysis are listed. For this, the articles were listed in order of ranking from the most cited to the least one, and in consequence of this, some developments in the area appear non-linearly in the text. A general analysis of the contributions in this entire set of articles is commented on as an overview.

The analysis begins with the literature review (\textbf{a01}) \cite{belanger2011privacy}, which focused on information systems and used data collected from students in the USA. Also, this work listed four dimensions to be safeguarded in the information age: privacy, accuracy, property, and accessibility $-$ all directly related to information security.

Next is the work \textbf{a02} \cite{xiao2012security} that brought an analysis of security and privacy in Cloud Computing by presenting its challenges and threats. In this work, threats were grouped according to five security attributes identified by the authors: confidentiality, integrity, availability, accountability, and privacy-preserving.

As the next article analyzed (\textbf{a03}) \cite{bhunia2014hardware} is a discussion of threat analysis and countermeasures for hardware Trojans. The authors presented this attack as an important security threat since it leads to a new set of reliability challenges of electronic equipment in the field $-$ also citing problems in military equipment that can be attributed to this type of attack.

In addition, on hardware security (\textbf{a04}) \cite{rostami2014primer}, a systematization of hardware security was presented that includes a classification of threats, state-of-art defenses, and evaluation of important metrics in hardware-based attacks.

Not so far from Hardware, Cyber-Physical Systems (CPS) in \textbf{a05} \cite{humayed2017cyber} presented a unified framework for CPS analysis. The work considered four areas of CPS: industrial control systems, smart grid systems, medical devices, and smart cars. Since, for each area listed, its threats and countermeasures existing in the state-of-art were analyzed. Furthermore, the analysis of each threat was based on five factors: origin, target, motive, attack vector, and potential consequences.

The \textbf{a06} \cite{carlini2016hidden} have analyzed aspects associated with hidden voice commands; their threats and possible mitigations are listed. Further, the voice is an open channel to proximity attacks, and some commands cannot be intelligible to humans; consequently, some types of defenses for these attacks are challenging. Thus, some defenses for hidden voice commands are categorized as notifications, challenges, or detection and prevention. However, active defenses can decrease the ease of use of devices leading to a decrease in their usability.

In \textbf{a07} \cite{cheminod2012review}, the object of study is industrial networks. The paper presented a review of the security perspective in the evaluation of the class of industrial distributed computing systems. Still, the authors outline four steps for controlling security risk prevention: definition of objectives, attack/threat models, security validation/analysis, and performance evaluation.

Research in \textbf{a08} \cite {bolle2002biometric} presented threats and mitigation in using physical characteristics (biometrics) in computer systems as a means of authentication. The work cites the Dodding zoo, adding another category of users - the chameleons - to whom there are open problems for authentication. Further, this reference addresses important issues regarding using biometrics as an authentication factor and false positives in some situations.

The trojans hardware theme is resumed for work \textbf{a09} \cite{xiao2016hardware}. Such a study was presented and analyzed a used adversary model and learned lessons in the last decade. Additionally, this work reinforces the challenge of dealing with this type of threat because intelligent adversaries design hardware Trojans to be undetectable.

Finally, there is \textbf{a10} \cite{yan2012security}, an work on towards in vehicular cloud computing. This type of computing intends to use the equipment on the side of highways and sources that have been idle for some time to create a cloud with a specific purpose. However, this type of computing presents challenges such as the authentication of sources with high mobility, scalability and a single interface, identity and location tangles, and the complexity of establishing trust relationships.

As perceived, the top ten works can be grouped into hardware security \cite{bhunia2014hardware,rostami2014primer,xiao2016hardware} with 30\% of the works, cloud computing \cite{xiao2012security,yan2012security} with 20\%, Cyber-Physical Systems \cite{humayed2017cyber,cheminod2012review} with 20\%, biometric factors \cite{carlini2016hidden,bolle2002biometric} with 20\% as well, and 10\% general purpose \cite{belanger2011privacy}. In this way, we can realize that one of the areas that most contribute and uses threat modeling in security and authentication in this study is engineering since it is responsible for the areas of hardware design and CSP.

\section{Results and Discussion}
\label{S:3}

While performing this bibliometric analysis, we achieved different objectives initially proposed through quantitative results obtained during the research. The first was to obtain a \textbf{representative dataset} of the selected area (the area of authentication and threat model) comprising the last 28 years of publications and a total of 922 documents. Therefore, some characteristics were analyzed from this dataset, as commented below.

Detailing the analysis of the dataset, we can list the results obtained in different items. In the \textbf{sources} item, the "Proceedings" obtained the highest number of publications among the types of sources (Table \ref{tab:most_relevant_sources}) with the highest number of citations. In the "Journal" type, the source Computer Networks (Elsevier) and the Internet of Things Journal (IEEE) were the best placed in terms of the number of articles.

Furthermore, as for publications, it was noted that the \textbf{highest publication volume} happened in the last ten years and had its \textbf{peak of average citations} in 2021 (very recently), showing the relevance and growth of this area nowadays. As a result, this area has generated new publications that have led to new and important research today.

\begin{figure}
    \centering
    \includegraphics[width=.75\textwidth]{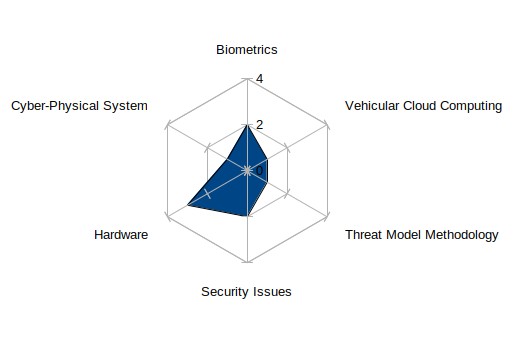}
    \caption{Top 10 Articles Categories}
    \label{fig:radar_categories}
\end{figure}

As for the articles, the \textbf{top ten most cited} from the dataset were brought and commented on in this study. So, Figure \ref{fig:radar_categories} was compiled from this list, which presents the occurrence of the proposed categorization (Biometrics, Vehicular Cloud Computing, Threat Model Methodology, Security Issues, and Hardware) for the most cited articles. We notice that the category with the most occurrences is Hardware, followed by biometrics and security issues, tied in second place. Based on this fact, we can conclude that hardware research has expressiveness and generates broadly new work, leading the area of authentication and threat models in our dataset.

Finally, we satisfactorily answered our initial questions (\textbf{Q1}, \textbf{Q2}, and \textbf{Q3}). Firstly, \textbf{Q1} is answered through the overviews presented in the \ref{S:2.2} subsection. Further, \textbf{Q2} is answered in Table \ref{tab:most_relevant_sources} and also commented in this section. Finally, \textbf{Q3} is presented in Table \ref{tab:top10_docs_citation} and categorized in Figure \ref{fig:radar_categories}, still in this same section.

\section{Conclusion and Future Work}
\label{S:4}

We can argue that our work fully achieved its proposal by analyzing variables and presenting an overview of threat model and authentication. Also, it was possible to present research trends within the studied area, the distribution pattern of publications over time, and discover the main publication sources, published works, and their authors. Therefore, the contributions of the ten articles with the most citations of this work were analyzed and commented on, allowing us to observe the documents that most contribute to the state-of-art in the desired themes. In addition, all questions (Q1, Q2, and Q3) have been answered satisfactorily.

As advances in the present study  is indicated an analysis of trends in the use of keywords, partnerships between authors (co-authorship), co-citation, and keywords co-occurrence, among other possible analyses. Although, all our goals have been met, it is possible to provide a depper overview of the area for future research by complementing the work through the analysis cited above and to a Forwarding Snow Balling process on the top ten articles.


\bibliographystyle{cas-model2-names}

\bibliography{cas-refs}

\end{document}